\documentclass{article}

\usepackage[accepted]{icml2021}
\usepackage[utf8]{inputenc} 
\usepackage[T1]{fontenc}    
\usepackage{hyperref}       
\usepackage{url}            
\usepackage{booktabs}       
\usepackage{amsfonts, amssymb, amsmath}       
\usepackage{xfrac}       
\usepackage[final]{microtype}      

\usepackage[most]{tcolorbox} 

\usepackage{tikz}

\usepackage[group-separator={,}]{siunitx}

\usepackage{graphicx}
\usepackage{subcaption}
\graphicspath{{figures/}}
\usepackage{wrapfig} 

\usepackage[capitalise]{cleveref} 

\newcommand{\cN}{\mathcal{N}}
\newcommand{\cK}{\mathcal{K}}
\newcommand{\cO}{\mathcal{O}}
\newcommand{\cP}{\mathcal{P}}

\newcommand{\cS}{\mathcal{S}}
\newcommand{\cA}{\mathcal{A}}

\usepackage[acronym]{glossaries-extra}
\glsdisablehyper
\setabbreviationstyle[acronym]{long-short}
\newacronym{ops}{SePS}{Selective Parameter Sharing}
\newacronym{nps}{NoPS}{No Parameter Sharing}
\newacronym{sap}{FuPS}{Full Parameter Sharing}
\newacronym{sapi}{FuPS+id}{Full Parameter Sharing with index}

\newcommand{\grad}[1]{\nabla #1} 

\DeclareMathOperator*{\dkl}{D_{KL}}

\usepackage{amsmath,amsfonts,bm}

\def\eqref#1{equation~\ref{#1}}

\def\1{\bm{1}}

\DeclareMathAlphabet{\mathsfit}{\encodingdefault}{\sfdefault}{m}{sl}
\SetMathAlphabet{\mathsfit}{bold}{\encodingdefault}{\sfdefault}{bx}{n}

\newcommand{\E}{\mathbb{E}}

\icmltitlerunning{Scaling Multi-Agent Reinforcement Learning with Selective Parameter Sharing}

\begin{document}

\twocolumn[
\icmltitle{Scaling Multi-Agent Reinforcement Learning with Selective Parameter Sharing}



\icmlsetsymbol{equal}{*}

\begin{icmlauthorlist}
\icmlauthor{Filippos Christianos}{ed}
\icmlauthor{Georgios Papoudakis}{ed}
\icmlauthor{Arrasy Rahman}{ed}
\icmlauthor{Stefano V. Albrecht}{ed}

\end{icmlauthorlist}

\icmlaffiliation{ed}{School of Informatics, University of Edinburgh, Edinburgh, United Kingdom}

\icmlcorrespondingauthor{Filippos Christianos}{f.christianos@ed.ac.uk}

\icmlkeywords{Machine Learning, ICML}

\vskip 0.3in
]

\printAffiliationsAndNotice{} 

\begin{abstract}

Sharing parameters in multi-agent deep reinforcement learning has played an essential role in allowing algorithms to scale to a large number of agents. Parameter sharing between agents significantly decreases the number of trainable parameters, shortening training times to tractable levels, and has been linked to more efficient learning. However, having all agents share the same parameters can also have a detrimental effect on learning. We demonstrate the impact of parameter sharing methods on training speed and converged returns, establishing that when applied indiscriminately, their effectiveness is highly dependent on the environment. We propose a novel method to automatically identify agents which may benefit from sharing parameters by partitioning them based on their abilities and goals. Our approach combines the increased sample efficiency of parameter sharing with the representational capacity of multiple independent networks to reduce training time and increase final returns.

\end{abstract}


\section{Introduction}

Multi-agent reinforcement learning (MARL) aims to jointly train multiple agents to solve a given task in a shared environment. Recent work has focused on novel techniques for experience sharing~\cite{christianos_shared_2020}, agent modelling~\cite{albrecht_autonomous_2018}, and communication between agents~\cite{rangwala_learning_2020, zhang_succinct_2020} to address the non-stationarity and multi-agent credit assignment problems~\cite{papoudakis_dealing_2019}. A problem that has received less attention to date is how to scale MARL algorithms to many agents, with typical numbers in previous works ranging between two and ten agents.

One common implementation technique to facilitate training with a larger number of agents is \emph{parameter sharing} (e.g.\ \cite{gupta_cooperative_2017}) whereby agents share some or all parameters in their policy networks. In the literature, parameter sharing is typically applied indiscriminately across all agents, which we call \emph{naive}. Naive parameter sharing has been effective primarily due to the similar (if not identical) observation and reward functions between agents found in many multi-agent environments. This similarity allows agents to share representations in intermediate neural network layers. Despite the occasional effectiveness of naive parameter sharing, it is not supported by theoretical work and has not received much attention beyond being mentioned as an implementation detail. Indeed, naive parameter sharing can decrease training time, but we show that it can be detrimental to final convergence in many environments, even when paired with implementation details that generally accompany it. We observe in our experiments that when the transition or the reward functions are distinct across agents, hidden representations that can be shared are harder to form, and fully-shared parameters are not effective.

\renewcommand{\thefootnote}{\fnsymbol{footnote}}
These limitations, however, do not imply that parameter sharing does not have a place in MARL. In contrast, we believe that it is an essential tool in scaling deep MARL algorithms to large numbers of agents, provided it can be done selectively, so as not to limit final performance. Therefore, we aim to benefit from parameter sharing when possible but also avoid potential bottlenecks. We introduce a method, \gls{ops}\footnote{We provide an open-source implementation of \gls{ops} here: \url{https://github.com/uoe-agents/seps}}, to automatically identify agents which may benefit from sharing parameters by partitioning them based on their abilities and goals. This partitioning is performed by encoding each agent to an embedding space by observing their trajectories, and then applying an unsupervised clustering algorithm to the encodings.

We can acquire an intuition of the setting this paper discusses by imagining a team of robots that must learn to run a restaurant, fulfilling both waiters and cooks' roles. Of course, agents belonging to the same group have to learn similar policies and therefore shared latent representations can significantly decrease learning requirements (i.e. there is no need for each cook to learn separately how to chop ingredients). Nevertheless, waiters and cooks have almost no common functionalities, and the representational capacity of a single neural network poses a bottleneck when attempting to learn all distinct roles. Furthermore, we show that agents tend to forget information needed by others when updating the parameters with their own objectives, actively interfering with other agents' learning.


We provide comparisons of typical usage of parameter sharing (sharing across all agents, appending agent indices, or not sharing at all) and show that i) \gls{ops} can converge to higher returns than both not sharing parameters and sharing them naively, ii) \gls{ops} is more sample efficient and executes considerably faster than not sharing parameters. Moreover, in contrast to baseline methods, \gls{ops} scaled to hundreds of agents (we experimented with up to 200) in our environments that contained non-homogeneous agents.


\section{Background}\label{sec:background}

\textbf{Markov Games:}
A Markov game (e.g. \cite{littman_markov_1994}; also known as a stochastic game~\cite{shapley1953stochastic}) with partial observability is defined by the tuple $(\cN,\cS, \{O^i\}_{i\in \cN}, \{A^i\}_{i\in \cN}, \cP, \{R^i\}_{i\in \cN})$, with agents $i\in\cN = \{1,\ldots,N\}$, state space $\cS$, joint observation space $\cO = O^1\times\ldots\times O^N$, and joint action space $\cA = A^1\times\ldots\times A^N$. Each agent $i$ only perceives local observations $o^i \in O^i$ which depend on the current state. Function $\cP: \cS \times \cA \mapsto \Delta(\cS)$ returns a distribution over successor states given a state and a joint action; $R^i: \cS \times \cA \times \cS \mapsto \mathbb{R}$ is the reward function giving agent $i$'s individual reward $r^i_t$ at timestep $t$. The objective is to find policies $\pi = (\pi_1, ..., \pi_N)$ for all agents such that the discounted return of each agent $i$, $G^i=\sum^T_{t=0}{\gamma^tr_t^i}$, is maximised with respect to other policies in $\pi$, formally $\forall_i : \pi_i \in \arg\max_{\pi'_i} \mathbb{E}[G^i | \pi'_i,\pi_{-i}]$ where $\pi_{-i} = \pi \setminus \{\pi_i\}$, $\gamma$ is the discount factor, and $T$ the total timesteps of an episode.

Unlike some recent MARL work we do not assume identical action, observation spaces, or reward functions between agents~\cite{christianos_shared_2020, rashid_qmix_2018, foerster_counterfactual_2018}.

\textbf{Policy Gradient and Actor-Critic: }
The goal of reinforcement learning is to find strategies that optimise the returns of the agents. Policy gradient, a class of model-free RL algorithms, directly learn and optimise a policy $\pi_\phi$ parameterised by $\phi$. The REINFORCE algorithm~\cite{williams_simple_1992}, follows the gradients of the objective $\grad_\phi J(\phi)=\E_{\pi_{\phi}}\left[G_t\grad_\phi{\ln{\pi_\phi(a_t|s_t)}}\right]$ to find a policy that maximises the returns. To further reduce the variance of gradient estimates, actor-critic algorithms replace the Monte Carlo returns with a value function $V_\pi(s; \upsilon)$. In a multi-agent, partially observable setting, a simple actor-critic algorithm defines the policy loss function for an agent $i$ as:
\begin{equation*}
    \label{eq:policyloss}
    \mathcal{L}(\phi_i) = -\log\pi(a_t^i|o_t^i;\phi_i)(r_t^i + \gamma V(o_{t+1}^i;\upsilon_i)-V(o_t^i;\upsilon_i))
\end{equation*}
and the respective value loss function as:
\begin{equation*}
    \label{eq:valueloss}
    \mathcal{L}(\upsilon_i) = ||V(o_t^i; \upsilon_i) - y_i||^2 \text{ \ with \ } y_i = r_t^i + \gamma V(o_{t+1}^i;\upsilon_i)
\end{equation*}

In this paper, for reinforcement learning, we use A2C~\cite{mnih_asynchronous_2016}, an actor-critic algorithm that additionally uses n-step rewards, environments that run in parallel, and improved exploration with entropy regularisation.

\textbf{Variational Autoencoders:} Variational autoencoders (VAEs) are generative models that explicitly learn a density function over some unobserved latent variables $Z$ given an input $x\in X$, where $X$ is a dataset. Given the unknown true posterior $p(z|x)$, VAEs approximate it with a parametric distribution $q_\theta(z|x)$ with parameters $\theta$. Computing the KL-divergence from the parametric distribution to the true posterior results in:
\begin{multline*}
    \dkl(q_\theta(z|x) \Vert p(z|x)) =  \log p(x) \\
     - \E_{z\sim q_\theta(z|x)}[\log p_u (x|z)] + \dkl(q_\theta(z|x) \Vert q(z))
\end{multline*}

The term $ \log p(x)$ is called log-evidence and it is constant. The other two terms are the negative evidence lower bound (ELBO). Minimising the ELBO is equivalent to minimising the KL-divergence between the parametric and the true posterior.

\section{Selective Parameter Sharing}\label{sec:methodology}

\begin{figure}
    \includegraphics[width=\linewidth]{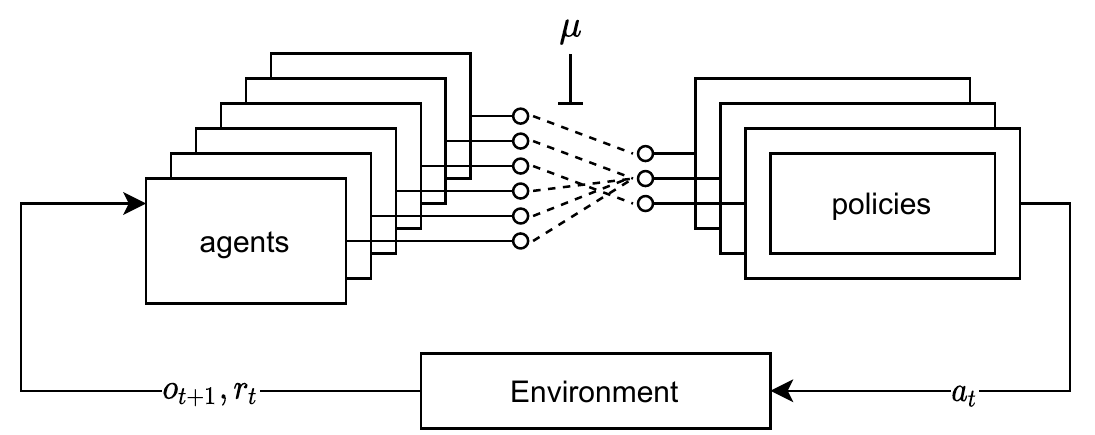}
    \caption{A top-level diagram of the selective parameter sharing architecture. $N$ agents are operating in one environment and receive observations and rewards. With \gls{ops} we are training $K < N$ policies and use a deterministic function $\mu$ to select which policy controls which agent.}
    \label{fig:diagram}
\end{figure}

To improve the effectiveness of parameter sharing, and allow for several distinct roles to be learned, we attempt to group agents that should be sharing their parameters during training. In an environment, we assume that $N$ agents can be partitioned into $K$ sets ($K<N$) but without knowing $K$ nor the partitioning. With $\cK=\left\{\pi_1,\ldots,\pi_K\right\}$, each agent in a cluster $k$ uses and updates the shared policy $\pi_k$. As we show in our experiments, such distinct shared policies can often be trained more efficiently while offering enough representational capacity to successfully solve the environment, and may even reach higher overall returns than alternative methods (see \cref{sec:exps}). \Cref{fig:diagram} depicts a top-level diagram of the components in our architecture.

To assign agents to partitions, we propose the use of a deterministic function $\mu:\cN\mapsto\cK$ that maps each agent $i$ to a parameterised policy (or partition) $\pi_k$. This partitioning is learned prior to RL training. Therefore, agents that share parameters get to benefit from shared representations in the latent layers of their neural networks, while not interfering with agents using other parameters. 


Recall the transition $\cP$ and reward $R^i$ functions that define an environment's dynamics (\cref{sec:background}). We aim to determine $\mu$ and partition agents such that agents which try to solve similar tasks use shared policies. Therefore, we introduce another concept: a set of functions $\hat{\cP^i}$ and $\hat{R^i}$ that attempt to approximate $\cP$ and $R^i$, but from the agents' limited perspective of the world. An agent does not observe the state nor the actions of another agent, and hence we define $\hat{\cP^i}: O^i\times A^i \mapsto \Delta(O^i)$ and $\hat{R^i}: O^i\times A^i \mapsto \mathbb{R}$, that model the next observation and reward respectively, based only on the observation and action of an agent $i$. When learning these functions our goal is not to ensure their accuracy as approximators of the dynamics; but rather that they identify similar agents to provide a basis for partitioning. We speculate that agents that should be grouped have similar reward and observation functions. Thus, the reasoning behind the following method is our desire to identify agents with identical $\hat{\cP^i}$ and $\hat{R^i}$, and have them share their network parameters.

\begin{figure}
    \includegraphics[width=\linewidth]{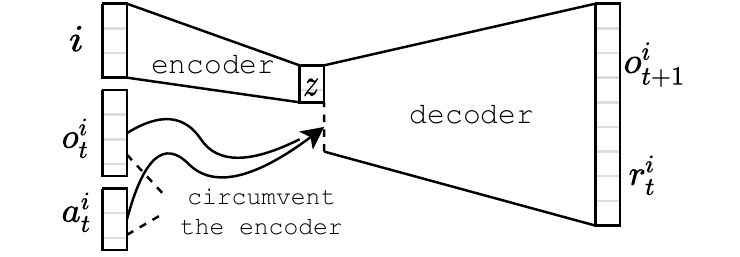}
    \caption{The encoder-decoder model. The encoder learns to encode the id of an agent in an embedding space while the decoder predicts the reward and next observation.}
    \label{fig:encoderdecoder}
\end{figure}

\subsection{Encoding Agent Identities}

We define an encoder $f_e$ and a decoder $f_p$ (\cref{fig:encoderdecoder}) parameterised by $\theta$ and $u$ respectively. The encoder, conditioned solely on the agent id, outputs the parameters that define an $m$-dimensional Gaussian distribution we can sample from.
We refer to samples from this latent space as $z$. The decoder, further divided into an observation and reward decoder $f^o_p$ and $f^r_p$ respectively, receives the observation, action, and sampled encoding $z$ of agent $i$, and attempts to predict the next observation and reward. In contrast to the classical definition of autoencoders, $o^i_t$ and $a^i_t$ \emph{bypass} the encoder and are only received by the decoder. Thus, due to the bottleneck, $z$ can only encode information about the agent, such as its reward function $\hat{R^i}$ or observation transition model $\hat{\cP^i}$.




To formalise the process, we assume that for each agent its identity $i$ is representative of its observation transition distribution and reward function. Additionally, we assume that both the identity of each agent and its observation transition distribution can be projected in a latent space $Z$ through the posteriors $q(z|i)$ and $p(z|\mathrm{tr} = (o_{t+1}, o_t, r_t, a_t))$. The goal is to find the posterior $q(z|i)$. We assume a variational family of parameterised Gaussian distributions with parameters $\theta$: $q_\theta(z|i) = \mathcal{N}(\mu_\theta, \Sigma_\theta; i)$. To solve this problem we use the variational autoencoding \citep{kingma_auto-encoding_2014} framework to optimise the objective $D_{KL}(q_\theta(z|i) || p(z|\mathrm{tr}))$.

We derive a lower bound on the log-evidence (ELBO) of the transition $\log p(tr)$ as:
\begin{multline}\label{eq:loss}
    \log p(\mathrm{tr}) 
    \geq \E_{z \sim q_\theta(z|i)}[\log p_u(\mathrm{tr}|z)] \\- D_{KL}(q_\theta(z|i) || p(z))
\end{multline}

The reconstruction term of  the ELBO factorises as:
\begin{equation*}
\begin{aligned}
    & \log p_u(\mathrm{tr}|z) =  \log p_u(r_t, o_{t+1}| a_t, o_t, z)p(a_t, o_t| z) = \\
    &\log p_u(r_t | o_{t+1}, a_t, o_t, z)  + \log p_u(o_{t+1} |a_t, o_t, z) + c
\end{aligned}
    \end{equation*}
    
The last term is discarded as $a_t$ and $o_t$ do not depend on the latent variable $z$. In these instances, $q_\theta$ and $p_u$ function as $f_e$ and $f_p$ respectively.

For the encoder-decoder model to learn from the experience of all agents, it is trained with samples from \emph{all} agents and will represent the collection of the agent-centred transition and reward functions $\hat{\cP^i}$ and $\hat{R^i}$ for all $i \in \cN$. Given the inputs of the decoder, the information of the agent id can only pass through the sample $z$. 


Minimising the model loss (\cref{eq:loss}) can be done prior to reinforcement learning. We sample actions $a^i\sim A^i$ and store the observed trajectories in a shared experience replay with all agents. We have empirically observed that the data required for this procedure is orders of magnitude less than what is usually required for reinforcement learning, and can even be reused for training the policies, thus not adding to the sample complexity.

The final step of the pre-training procedure is to run a clustering algorithm on the means generated from the encoder $f_e(i)$ for all $i\in N$, and use the agent indices clustered together to define $\mu$. In the experiments that follow, we use k-means for simplicity. After the partitioning is completed, a static computational graph (for automatic differentiation) can be generated to train the policies with significant speed advantages.

\section{Experimental Evaluation}\label{sec:exps}

In this section, we evaluate both whether \gls{ops} performs as intended by correctly partitioning the agents and whether this partitioning helps in improving the overall returns, sample complexity, and training time.  For RL, we use the A2C~\cite{mnih_asynchronous_2016} algorithm and report the sum of returns of all agents. A search was performed for A2C's hyperparameters across all baselines, while hyperparameters of the clustering portion of \gls{ops} were easily found manually, and kept identical across all environments (more details in \cref{sec:reproducibility}).

\subsection{Multi-Agent Environments}\label{sec:environments}

\begin{figure*}[t]
     \centering
     \begin{subfigure}[t]{0.24\textwidth}
         \captionsetup{justification=centering}
         \centering
            \includegraphics[width=\textwidth]{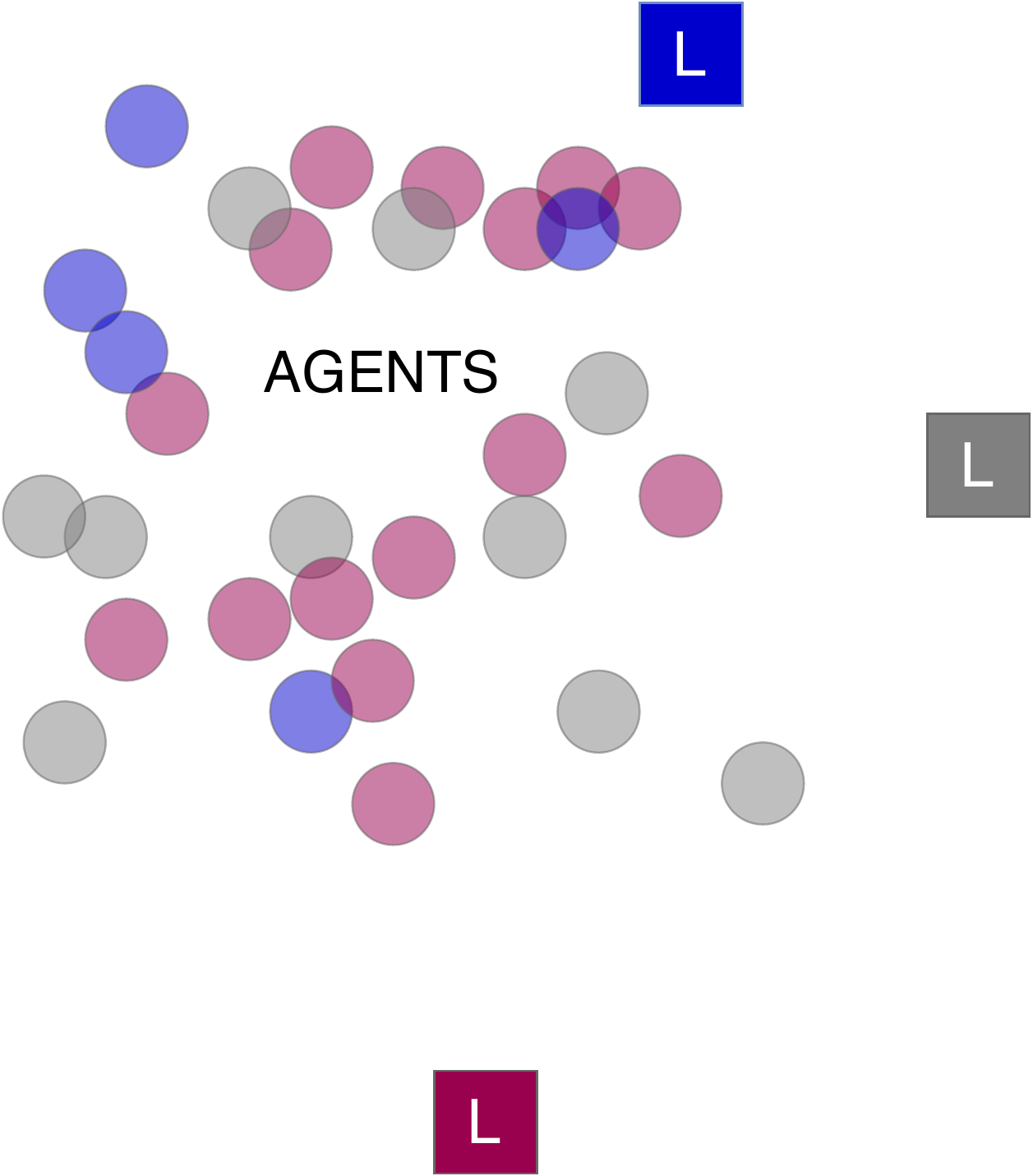}
         \caption{Blind-Particle Spread\\(BPS)}
         \label{fig:envs:blindspread}
     \end{subfigure}
     \hfill
     \begin{subfigure}[t]{0.24\textwidth}
         \captionsetup{justification=centering}
         \centering
            \includegraphics[width=\textwidth]{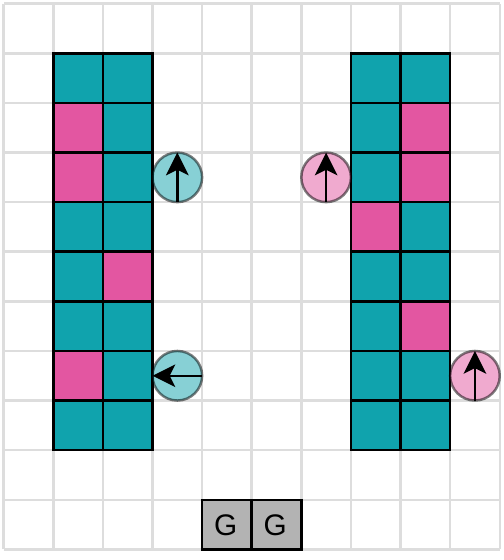}
         \caption{Coloured Multi-Robot\\Warehouse (C-RWARE)}
         \label{fig:envs:crware}
     \end{subfigure}
     \hfill     
     \begin{subfigure}[t]{0.24\textwidth}
         \captionsetup{justification=centering}
         \centering
            \includegraphics[width=\textwidth]{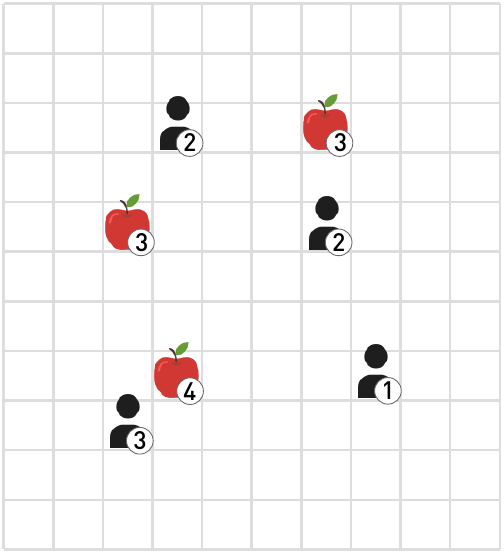}
         \caption{Level-based Foraging\\(LBF)}\label{fig:envs:lbf}
         \label{fig:lbf}
     \end{subfigure}
     \hfill
     \begin{subfigure}[t]{0.24\textwidth}
         \captionsetup{justification=centering}
         \centering
            \includegraphics[width=\textwidth]{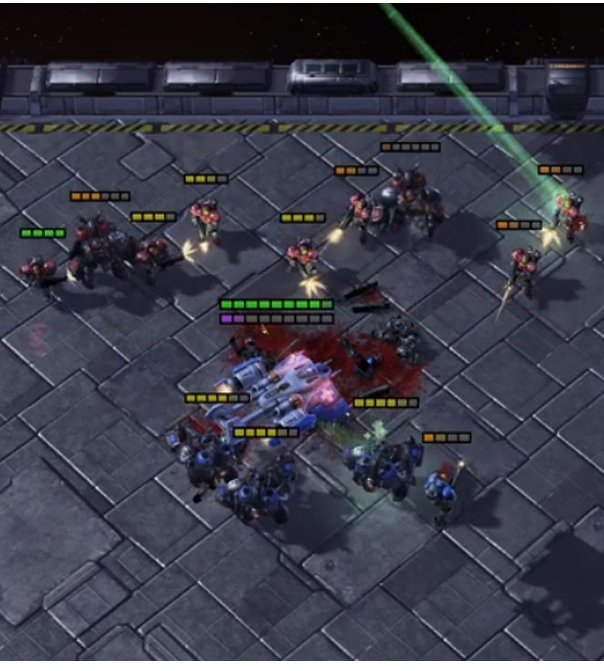}
         \caption{StarCraft Multi-Agent\\Challenge (SMAC/MMM2)}
         \label{fig:envs:smac}
     \end{subfigure}
    \caption{Visualisation of environments used in experiments.}
    \label{fig:envs}
\end{figure*}

We use four multi-agent environments (\cref{fig:envs}) which are described below and summarised in \cref{tab:brieftasks}.

\textbf{Blind-particle Spread:} Our motivating toy environment is a custom scenario created with the Multi-agent Particle Environment (MPE)~\cite{lowe_multi-agent_2017}. Blind-particle spread (BPS, \cref{fig:envs:blindspread}) consists of landmarks of different colours and numerous agents that have been also assigned colours. The agents are unable to see their own colour (or of the other agents) but need to move towards the correct landmark. This environment enables us to investigate the effects of parameter sharing by allowing us to control two important variables: i) the number of agents and ii) the number of colours (distinct behaviours that must be learned). In a further, more difficult variation which we name BPS-h, each group of agents also has a different observation space (e.g.\ the agents could be equipped with different sensors).  

\begin{table}[t]\centering
\caption{\label{tab:brieftasks}Brief description of environments, including how agents are distributed to different agent types: colours in BPS and C-RWARE, levels in LBF or different units in MMM2. Environments with † or ‡ mark different observation spaces or action spaces across types respectively, while § marks a cooperative (shared) reward. }
\begin{tabular}{@{}lrrr@{}}
\toprule
{}          & \# Agents & \# Types              & Type Distribution\\ \midrule
BPS (1)     & 15        & 3                     & 5--5--5            \\
BPS (2)     & 30        & 3                     & 10--10--10         \\
BPS (3)     & 30        & 5                     & 6--6--6--6--6      \\
BPS (4)     & 30        & 5                     & 2--2--2--15--9     \\
BPS-h (1)   & 15        & 3\textsuperscript{†}  & 5--5--5            \\ 
BPS-h (2)   & 30        & 5\textsuperscript{†}  & 6--6--6--6--6      \\ 
BPS-h (3)   & 200       & 4\textsuperscript{†}  & 50--50--50--50     \\ 
C-RWARE (1) & 4         & 2\textsuperscript{‡}  & 2--2               \\ 
C-RWARE (2) & 8         & 2\textsuperscript{‡}  & 4--4               \\ 
C-RWARE (3) & 16        & 2\textsuperscript{‡}  & 8--8               \\ 
LBF         & 12        & 3                     & 4--4--4--4         \\ 
MMM2        & 10\textsuperscript{§}        & 3  & 7--2--1            \\ \bottomrule
\end{tabular}
\end{table}

\textbf{Coloured Multi-Robot Warehouse:} The Coloured Multi-Robot Warehouse (C-RWARE, \cref{fig:envs:crware}) is a variation of the RWARE environment~\cite{christianos_shared_2020}, where multiple robots have different functionalities and are rewarded only for delivering specific shelves (denoted by different colours) and have different action spaces. The agents can rotate or move forward and pick up or drop a shelf. The observation consists only of a $3\times3$ square centred around the agent. Agents are only rewarded (with $1.0$) when successfully arriving at the goal with a requested shelf of the correct colour, making the reward sparse. RWARE is known~\cite{christianos_shared_2020, papoudakis_comparative_2020} to be an environment with difficult exploration, and independent learners have been shown to struggle on it. 

\textbf{Level-based Foraging:} Level-based Foraging (LBF, \cref{fig:lbf})~\cite{albrecht_game-theoretic_2013} is a multi-agent environment where agents are placed in a grid, and required to forage randomly scattered food. Each agent is assigned a level, and each food also is assigned a level at the beginning of the episode. The agents can move in four directions and attempt to forage an adjacent food. For foraging to be successful, the sum of the agent levels foraging the food must be equal or greater than its level. LBF is partially observable, and while the agents can see the positioning of agents and food, as well as the food levels, they cannot see any of the agent levels. The reward is proportionate to the agents' contribution when a food is successfully loaded.

\textbf{Starcraft Multi-Agent Challenge:} While the multi-agent Starcraft (SMAC)~\cite{samvelyan_starcraft_2019} environment might not be the archetype for displaying the strengths of selective parameter sharing, it is a widely used setting where multiple agents of distinct types co-exist and must learn together. For instance, the ``MMM2'' environment (\cref{fig:envs:smac}) contains three types of units (marines, marauders, and medivacs) with distinct attributes. One of those unit types, medivacs, is especially different since it needs to learn how to heal friendly units instead of attacking enemies.


\subsection{Baselines}

We compare \gls{ops} against several other methods of parameter sharing described below.

\textbf{\gls{nps}:} In our \gls{nps} baseline, all agents have their own set of parameters, and there is no overlap of gradients. This approach is common in the literature and usually encountered when there is no mention of parameter sharing, e.g.\ MADDPG~\cite{lowe_multi-agent_2017}.

\textbf{\gls{sap}:} The second baseline, \gls{sap} consists of a single set of parameters that will be shared between all agents. \Gls{sap} is a naive baseline since it does not allow agents to ever develop any differences in their behaviour.

\textbf{\gls{sapi}:} Finally, we test a variation of the previous method, where the policy is also conditioned on the agent id. While the use of \gls{sap} is limited and our expectations are not high, since there is no way to differentiate between agents, \gls{sapi} is encountered very often in the literature~\cite{rashid_qmix_2018,foerster_counterfactual_2018,gupta_cooperative_2017}. 

\subsection{An Experimental Evaluation of \glsfmtshort{sapi}}
\label{sec:exps:sap-i}
\begin{figure}[t]
    \centering
    \includegraphics[trim=0 0 0 40,clip,width=\linewidth]{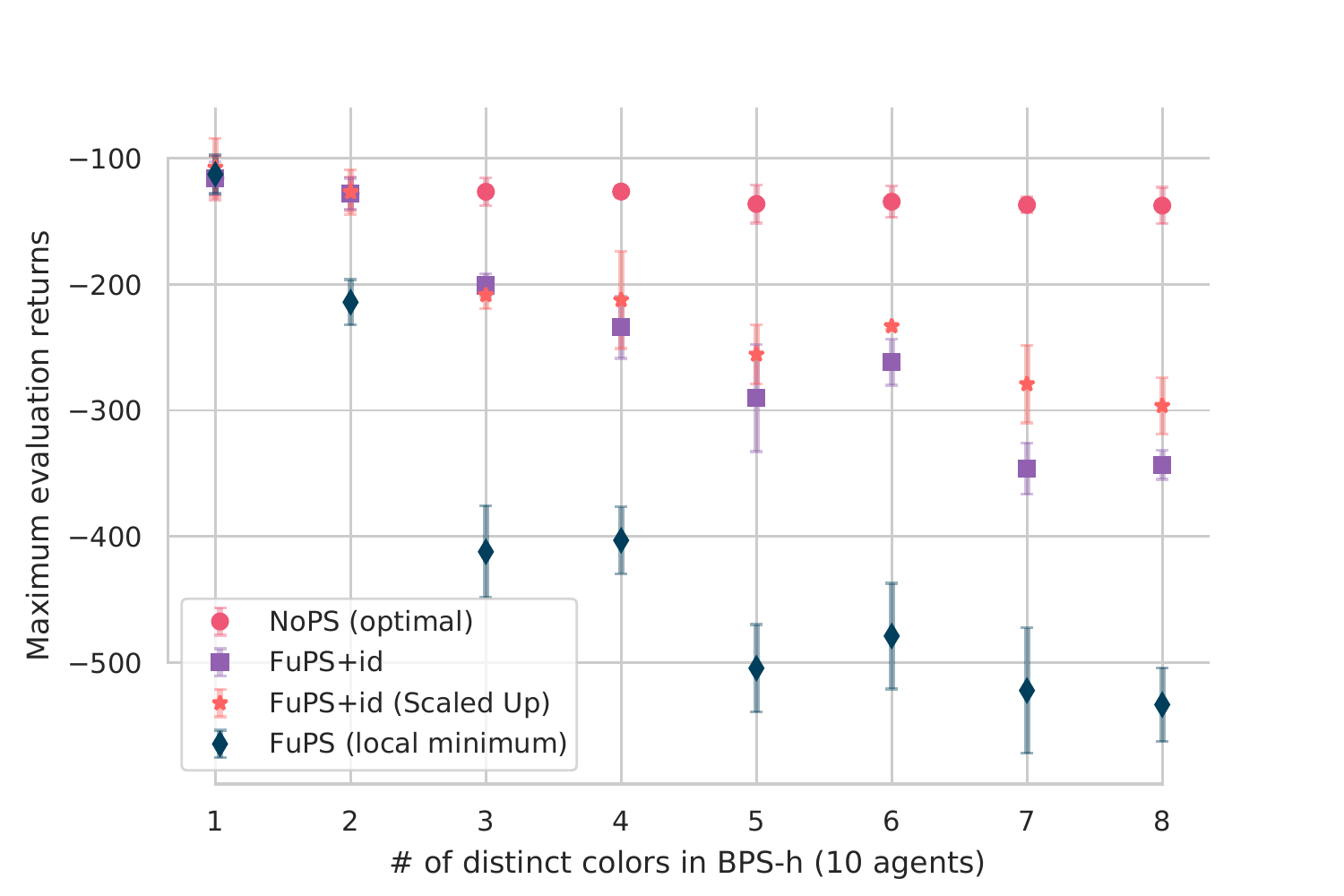}
    \caption{BPS-h with 10 agents and a variable number of colours. The maximum evaluation returns are recorded for each environment and algorithm, and vertical bars indicate standard deviation across seeds.}
    \label{fig:colorprogression}
\end{figure}

The dense reward signal in our toy environment, BPS, makes it sufficiently simple for non-parameter sharing agents to learn how to reach their respective landmarks. However, when parameter sharing is involved, we expect the task to become considerably harder. \gls{sapi} presumably solves this by allowing each agent to develop a distinct policy based on agent id. To investigate this, we tested \gls{nps}, \gls{sap}, and \gls{sapi}, in a series of BPS tasks, where the number of agents will remain constant, and the number of colours (landmarks) increases. \gls{nps} should have no issue learning immediately since each agent needs to learn to navigate to a specific landmark. In contrast, sharing parameters with \gls{sap} can not work because the agents lack the information required to determine their colour and move accordingly. Therefore, agents trained with \gls{nps} tend to fall into the \emph{local minimum} of moving to a location that minimises the distance between all landmarks. 

We are, however, very interested in what \gls{sapi} can learn. The agents have all the information needed to learn how to move to the correct landmark. But, as we have hypothesised in earlier sections, the overlap of different policies which must be represented on the same parameters, poses a significant bottleneck for learning. Indeed, as \cref{fig:colorprogression} indicates, the performance of \gls{sapi} deteriorates sharply, even with only three colours.

An argument could be made that an increased number of colours in the BPS task should be accompanied by an increase in \gls{sapi}'s model size. Such an increase in the number of parameters increases the representation capacity and could allow for multiple distinct policies to be learned. To test this hypothesis we include in \cref{fig:colorprogression} the \gls{sapi} (Scaled Up) baseline which scales the width of the network such that the number of parameters grows with the number of colours shown in the $x$ axis. 
Specifically, in this experiment the \gls{sap} and \gls{sapi} baselines use approximately $18K$ parameters (two layers of 128 units). The \gls{sapi} (Scaled Up) baseline, however, uses approximately $\#Colors * 18K$ parameters for the different BPS-h tasks (for exact sizes of the layers in each task see \cref{sec:reproducibility}). 

However, we observe that even the \gls{sapi} (Scaled Up) baseline does not successfully learn these otherwise simple BPS tasks. Therefore we conjecture that the issue with \gls{sapi} is not the model capacity since it could have enough parameters to learn all behaviours. Instead, learning on shared parameters interferes with the learning of other agents. In the following sections, we will instead optimise the network size as a hyperparameter.

\subsection{Reinforcement Learning with Parameter Sharing}

\begin{figure*}[t]
     \centering
     \begin{subfigure}[b]{0.33\textwidth}
         \centering
            \includegraphics[width=\linewidth]{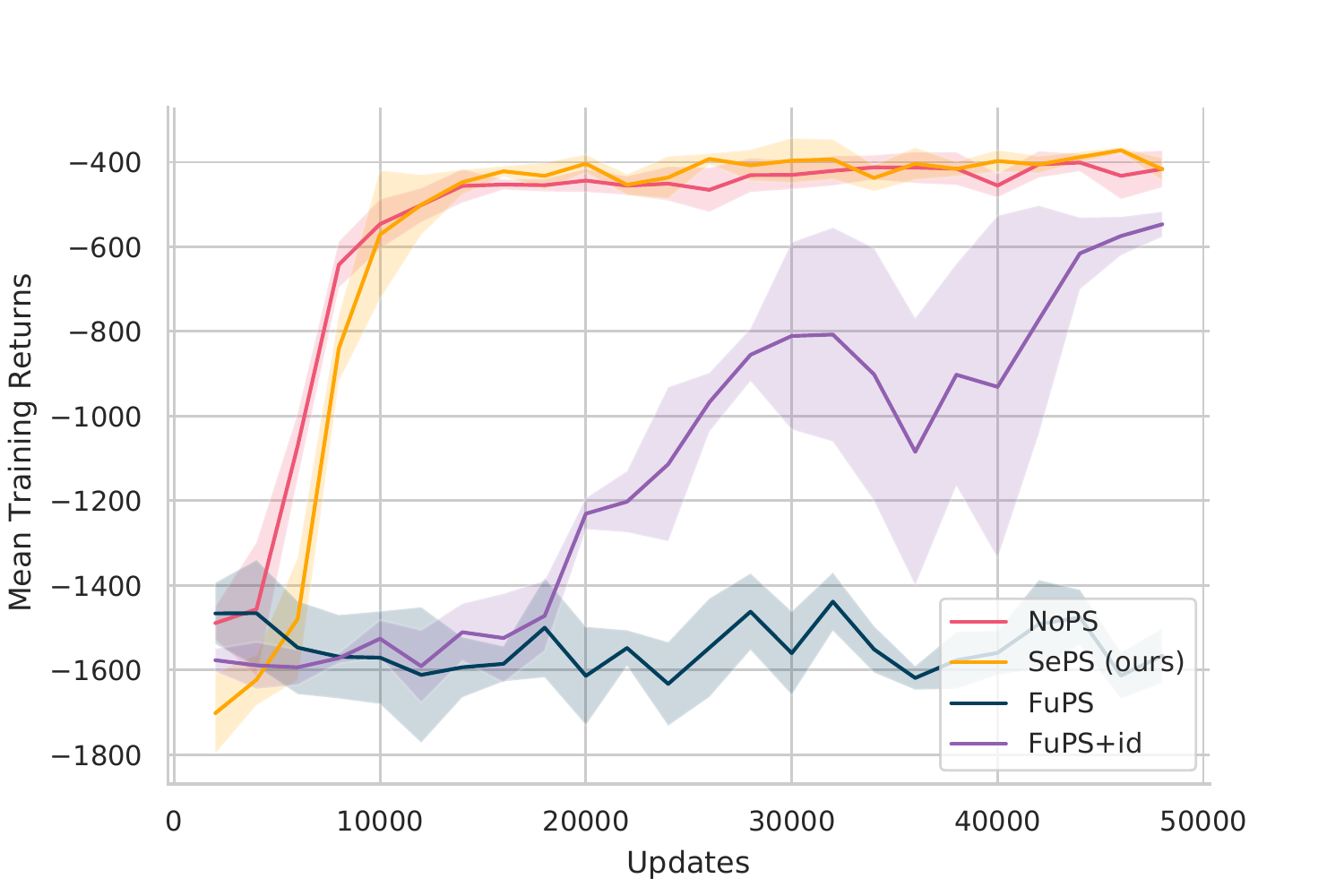}
         \caption{\label{fig:results:bps}BPS (3)}
     \end{subfigure}
     \hfill
     \begin{subfigure}[b]{0.33\textwidth}
         \centering
            \includegraphics[width=\linewidth]{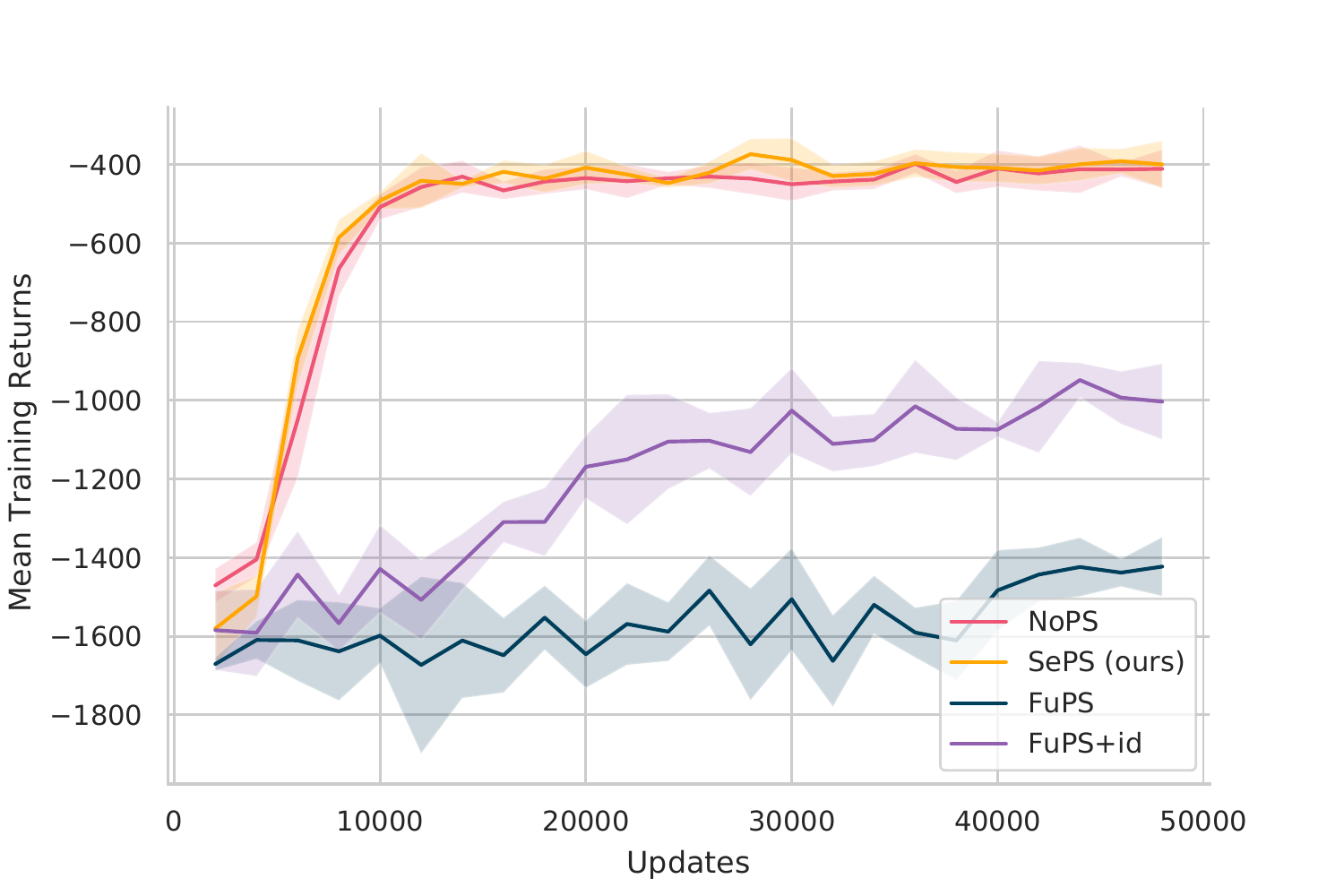}
         \caption{\label{fig:results:bpsh}BPS-h (2)}
     \end{subfigure}     
     \begin{subfigure}[b]{0.33\textwidth}
         \centering
            \includegraphics[width=\linewidth]{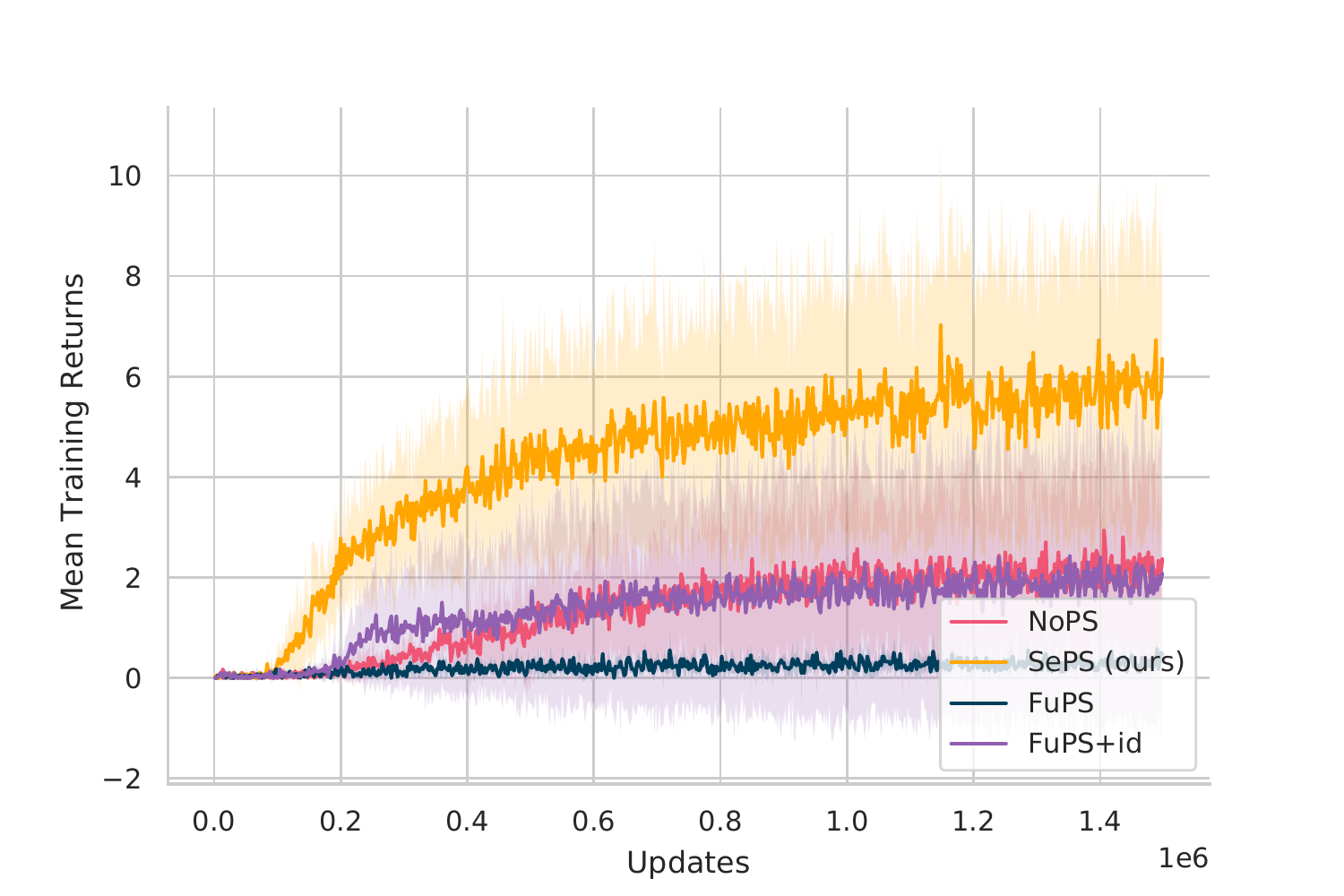}
         \caption{\label{fig:results:rware1}C-RWARE (1)}
     \end{subfigure}
      \begin{subfigure}[b]{0.33\textwidth}
         \centering
            \includegraphics[width=\linewidth]{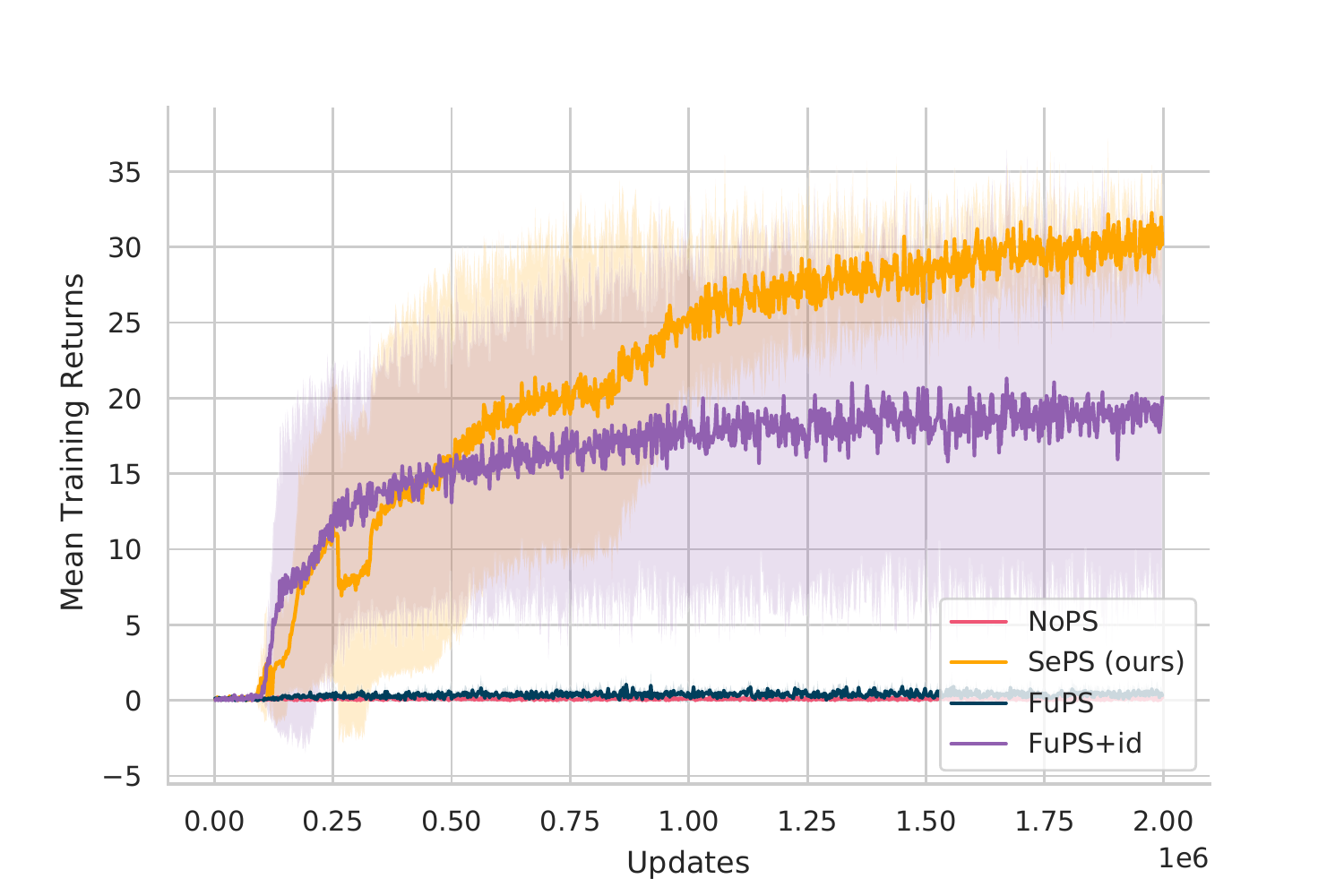}
         \caption{\label{fig:results:rware2}C-RWARE (3)}
     \end{subfigure}
     \begin{subfigure}[b]{0.33\textwidth}
         \centering
            \includegraphics[width=\linewidth]{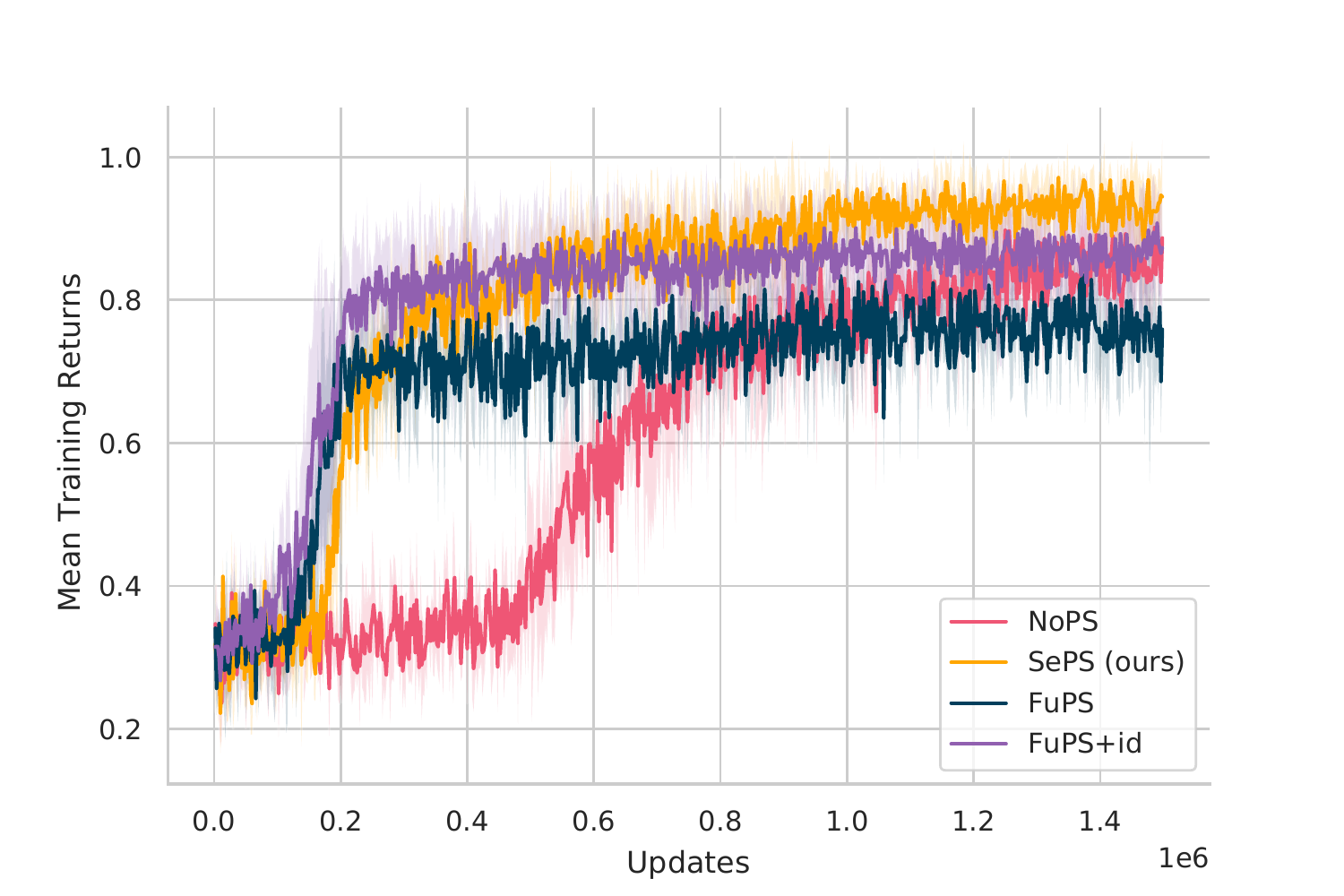}
         \caption{\label{fig:results:lbf}LBF}
     \end{subfigure}
      \begin{subfigure}[b]{0.33\textwidth}
         \centering
            \includegraphics[width=\linewidth]{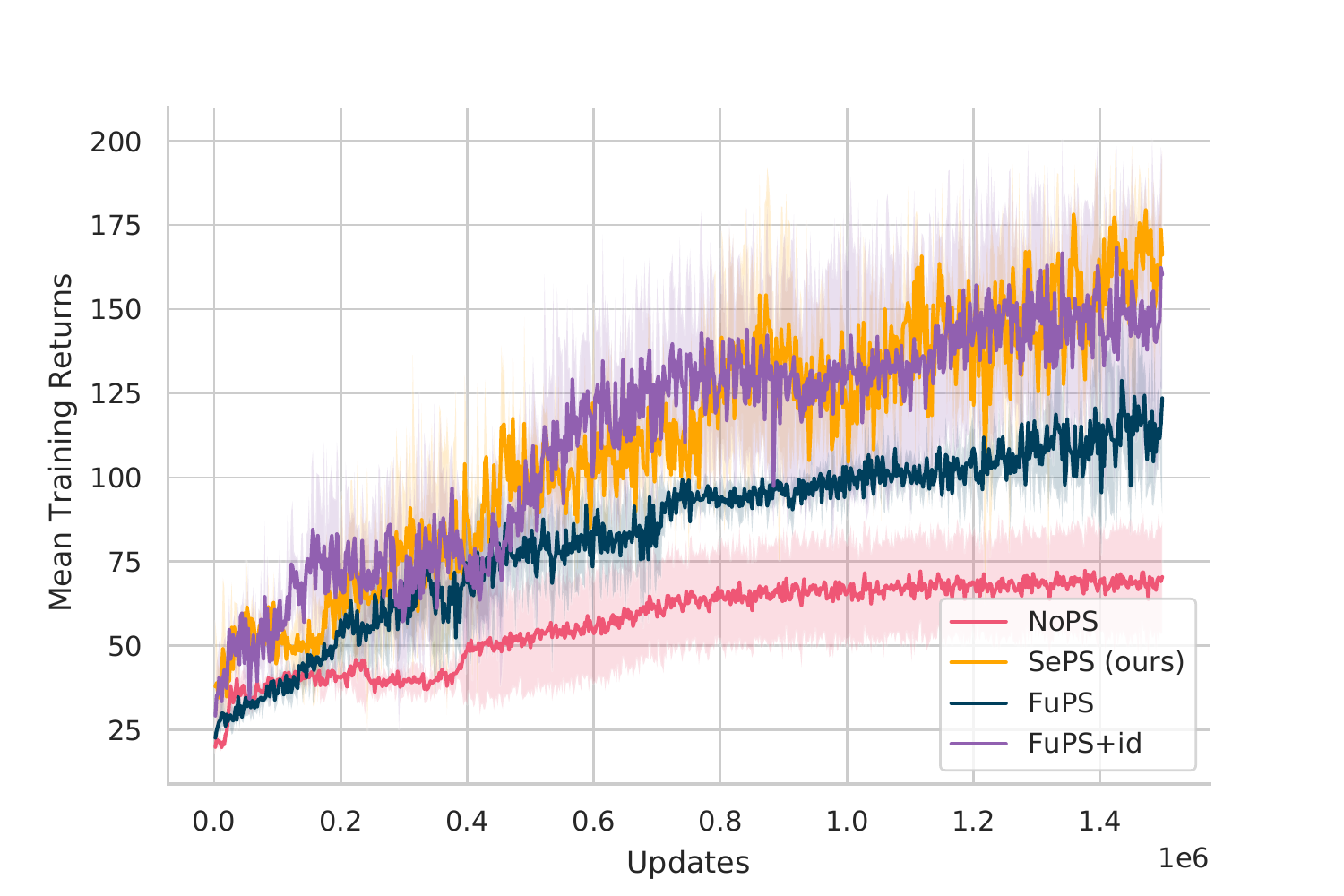}
         \caption{\label{fig:results:smac}SMAC (MMM2)}
     \end{subfigure}
    \caption{\label{fig:results}Learning curves showing the mean returns during training for a selection of the environments. The shaded area represents standard deviation across seeds.}
\end{figure*}

Next, we investigate how selectively sharing parameters affects reinforcement learning performance. Our hypothesis remains that agents can benefit from sharing parameters if they have been clustered together by \gls{ops}. Our results, detailed in \cref{tab:results} support this hypothesis. We also present the learning curves on a selection of environments in \cref{fig:results}.

\begin{table*}[t]\centering
\caption{\label{tab:results}Maximum evaluation returns with std across seeds. Highest means (within one std) are shown in bold.}
\begin{tabular}{lrrrr}
\toprule
{} &            \gls{nps} &     \gls{ops} (Ours) &              \gls{sap} &            \gls{sapi} \\
\midrule
BPS (1)     &    $\mathbf{-189.99 \pm 12.96}$ &     $\mathbf{-179.69 \pm 7.02}$ &    $-612.18 \pm 52.14$ &    $-221.44 \pm 9.72$ \\
BPS (2)     &    $\mathbf{-385.18 \pm 34.91}$ &     $\mathbf{-369.69 \pm 38.70}$ &  $-1301.35 \pm 164.84$ &   $-443.84 \pm 20.74$ \\
BPS (3)     &    $-401.09 \pm 20.37$ &     $\mathbf{-371.99 \pm 6.92}$ &   $-1438.41 \pm 68.41$ &   $-546.95 \pm 29.52$ \\
BPS (4)     &    $\mathbf{-388.21 \pm 41.35}$ &     $\mathbf{-378.50 \pm 77.75}$ &   $-1339.14 \pm 43.95$ &  $-722.33 \pm 103.92$ \\
BPS-h (1)   &    $\mathbf{-187.92 \pm 15.78}$ &     $\mathbf{-189.93 \pm 31.29}$ &    $-571.46 \pm 52.12$ &   $-293.31 \pm 26.89$ \\
BPS-h (2)   &    $\mathbf{-398.11 \pm 23.64}$ &     $\mathbf{-373.92 \pm 39.19}$ &   $-1422.80 \pm 74.70$ &   $-948.44 \pm 43.67$ \\
BPS-h (3)   &                   N/A  &  $\mathbf{-2522.61 \pm 276.29}$ &  $-6825.47 \pm 115.74$ &   $-4085.51 \pm 3.71$ \\
C-RWARE (1) &        $2.93 \pm 2.25$ &        $\mathbf{7.03 \pm 3.72}$ &        $0.57 \pm 0.28$ &       $2.42 \pm 3.52$ \\
C-RWARE (2) &        $0.28 \pm 0.28$ &       $\mathbf{20.88 \pm 1.15}$ &        $0.55 \pm 0.40$ &      $10.35 \pm 8.65$ \\
C-RWARE (3) &        $0.33 \pm 0.15$ &       $\mathbf{32.27 \pm 3.16}$ &        $1.03 \pm 0.64$ &     $\mathbf{21.30 \pm 15.13}$ \\
LBF         &        $0.91 \pm 0.05$ &        $\mathbf{0.97 \pm 0.03}$ &        $0.83 \pm 0.02$ &       $0.91 \pm 0.05$ \\
MMM2        &      $72.32 \pm 15.00$ &      $\mathbf{179.45 \pm 7.11}$ &     $128.75 \pm 18.37$ &    $\mathbf{168.39 \pm 29.99}$ \\

\bottomrule
\end{tabular}
\end{table*}

\textbf{BPS:} The BPS tasks (\cref{tab:results,fig:results:bps,fig:results:bpsh}) are trivial for independent learners given the dense reward: each agent learns to always move towards a specific colour. However, if all agents share a policy, then the agents (not being able to perceive their own colours) only learn to move to a local minimum. \gls{sapi}, which supposedly circumvents the problem, still has issues correctly learning this problem. Due to the high computational requirements of \gls{nps}, which requires $N$ different sets of parameters, running it on BPS-h (4) with 200 agents was infeasible. 

\textbf{C-RWARE:} Our results in C-RWARE (\cref{tab:results,fig:results:rware1,fig:results:rware2}) are more surprising. \gls{nps} which was a strong contender in BPS, completely failed to learn in C-RWARE (2) and (3). These tasks are very sparsely rewarded, which seems to make independent learning ineffective. Instead, sharing parameters also combined the received rewards, providing a useful learning direction to the optimiser. Also, similarly to BPS, \gls{ops} outperforms the other naive parameter sharing methods.

\textbf{LBF:} Similarly to other environments, \gls{ops} agents in LBF achieve higher returns with more efficient use of environment samples (\cref{fig:results:lbf}). The optimal performance in LBF is $1.0$, and while \gls{sapi} is close, it does not achieve the same returns as \gls{ops}. \Gls{nps} takes considerably more samples to train, but given our BPS results, it is possible it eventually converges to the same returns as \gls{ops}.

\textbf{MMM2: } In one of the hardest SMAC environments, MMM2 (\cref{fig:results:smac}), the most surprising result was the difference in converged returns between \gls{nps} and parameter sharing methods. Even fully shared parameters - and after making sure the identity of the agents does not leak through the observation - outperforms \gls{nps}. Our hypothesis on these results is that i) this task requires agents to act in a very similar way (e.g.\ only targeting the same opponents) and ii) parameter sharing plays a previously underrated role in decomposing (or reasoning over) a shared reward. The rest of the methods behave similarly to other environments, but with a minimal improvement of \gls{ops} over \gls{sapi}.

\subsection{A Peek into the Embedding Space of \glsfmtshort{ops}}

\begin{figure}[t]
    \centering
    \includegraphics[width=\linewidth,trim={2cm 2cm 2cm 2cm},clip]{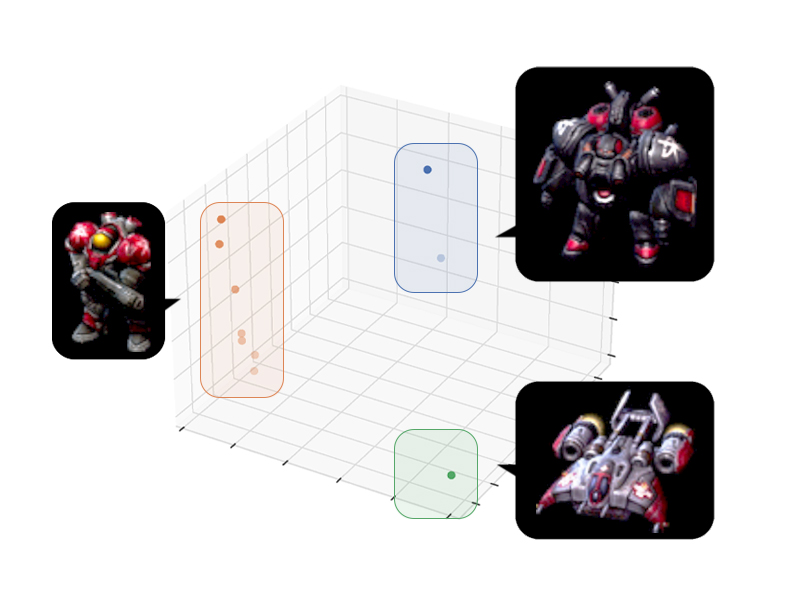}
    \caption{Visualisation of the means of z for each of the 10 agents in SMAC (MMM2 task) in a 3-dimensional space. The colours identify the clusters created by k-means.}
    \label{fig:sampled_z}
\end{figure}

Next, an important step in evaluating \gls{ops} is to verify that meaningful clusters appear after optimising the objectives discussed in \cref{sec:methodology}.

Our goal is to compare the clustering of our algorithm, with one decided by a human who is given knowledge of the environment. We visualise the embedding for each of the units on a SMAC task (\cref{fig:sampled_z}). We can assert that clearly visible clustering is precisely what we would expect. Different types of units all have different properties (e.g.\ movement speed, health, or damage) and thus a distinct interaction with the environment. These differences were picked up by the encoder that subsequently spread them in the embedding space.

A question that might arise is why the agents of the same unit type (and cluster) are spread out in the $z$ axis of \cref{fig:sampled_z}. In the SMAC environment, there is another difference between the agents: their starting position. Therefore, the initial observations $o^i_0$ are sampled from different sets for each of the agents. The encoder picks up on this feature and further spreads the latent encoding. While k-means clustered the agents by unit type, it could be argued that more clusters could have been formed, which goes beyond the scope of our work and is considered a feature (and open problem) of unsupervised clustering~\cite{kaufman_finding_2009}. However, this was the exception in our tests, and in all other environments, where the starting observations are sampled from the same set, the latent variables of similar agents are overlapping one another. Nevertheless, in \cref{sec:nclusters} we explore how the number of clusters can be determined, and how wrong choices can affect learning. 

The clustering process across all environments and seeds matched the various types of agents. For instance, in C-RWARE and BPS, each cluster contains only agents of the same colour. Importantly, this information is not included in the observation space and therefore not observed by the agents or even the encoder; it is only understood after observing the transitions and rewards of each agent.

\subsection{Determining the Number of Clusters}\label{sec:nclusters}

Several ways to determine the number of clusters in the embedding space exist. Arguably the most straightforward is the use of domain knowledge. But, having an estimate of the value of $K$, could also mean knowledge of which agents should be assigned in clusters in the first place. Despite the diminished importance of the pretraining \gls{ops} stage in this situation, we believe that understanding the effectiveness of shared parameters between clusters is still of value.

\begin{figure}[t]
    \centering
    \includegraphics[width=\linewidth]{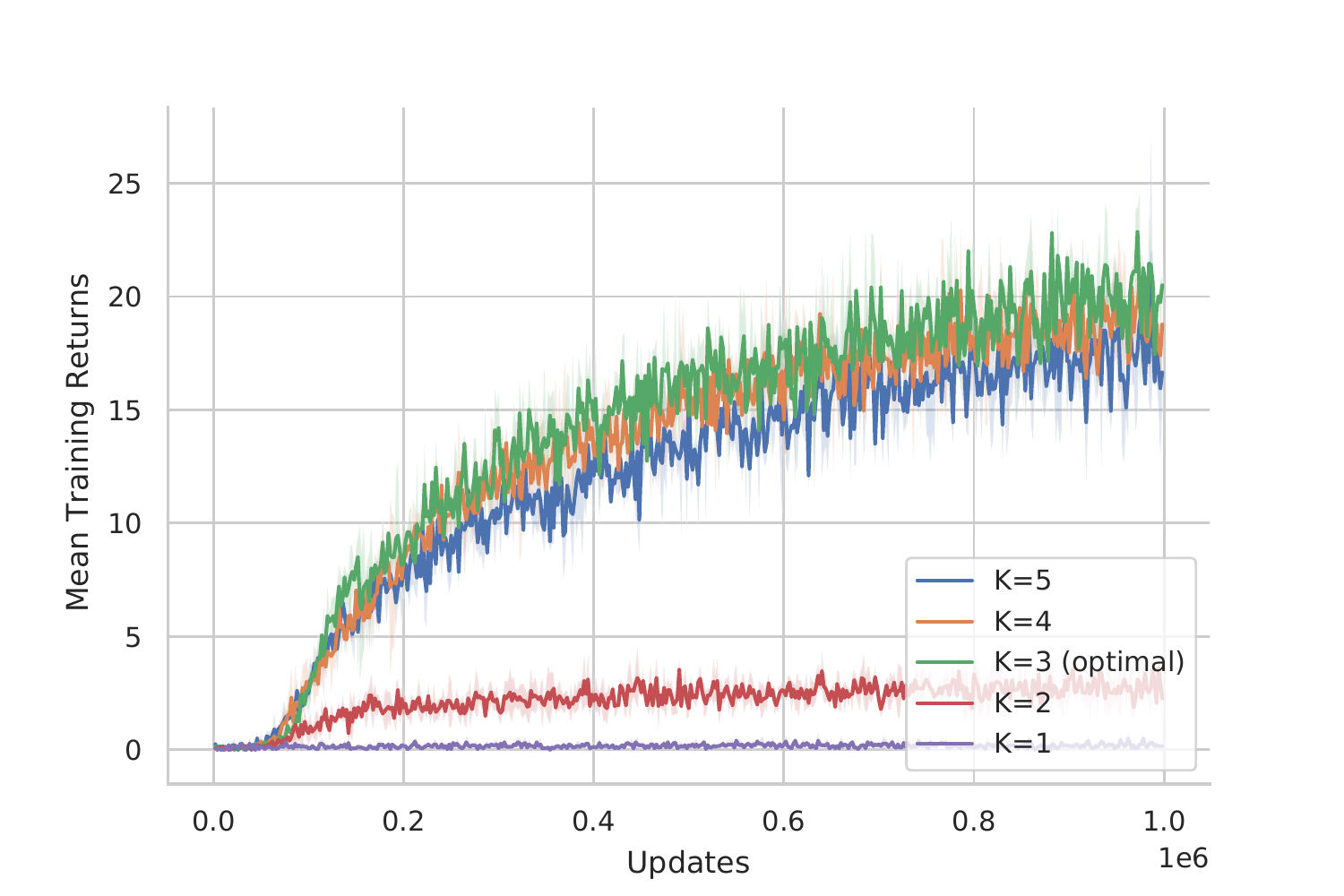}
    \caption{Mean returns during training for different number of clusters on C-RWARE (9 agents and 3 colours).}
    \label{fig:hyperk}
\end{figure}

A second approach would consist of treating $K$ as a tunable hyperparameter. In \cref{fig:hyperk}, we present the returns during training on C-RWARE when \gls{ops} is forced to create a varied amount of clusters. It is clear from the results, that overestimating $K$ is of little significance. However, trying to form fewer clusters than needed lowers the achieved returns, and collapses to \gls{nps} when $K=1$.

Finally, there is a plethora of well-studied heuristics for separating clusters when the embedding space is known. The elbow method~\cite{thorndike_who_1953}, the silhouette method~\cite{rousseeuw_silhouettes_1987}, or the Davies–Bouldin index~\cite{davies_cluster_1979}, all could be used to determine the number of clusters since our method tends to produce well-separated values. We have implemented and tested the Davies-Bouldin index, and we have found that coupled with k-means, reliably finds the same clusters an expert would in our tested environments (i.e. the second column in \cref{tab:brieftasks}).

\subsection{Computational Benefits}

\begin{figure}[t]
    \centering
    \includegraphics[width=\linewidth]{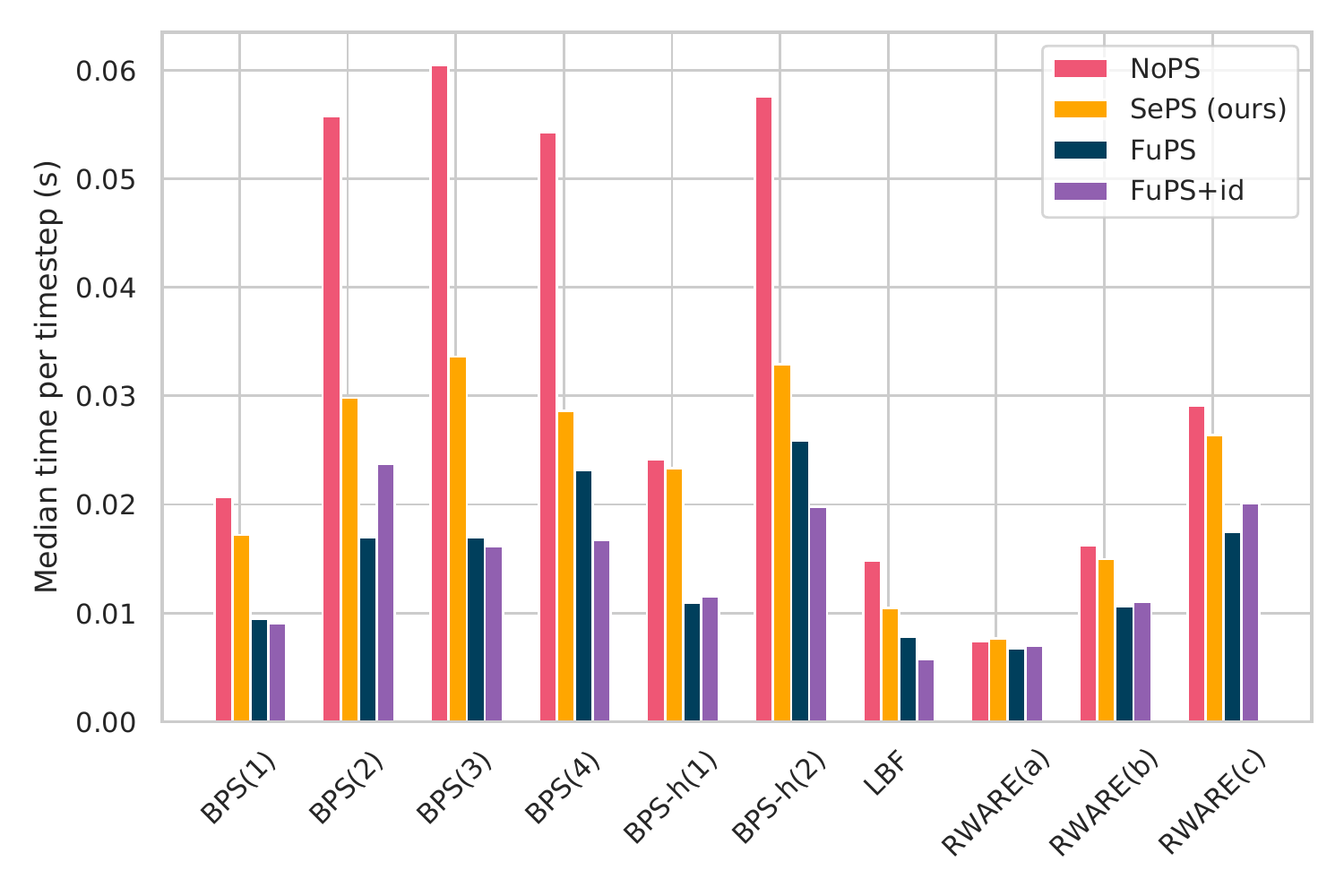}
    \caption{Median running time of a timestep during training over all the the environments and methods.}
    \label{fig:run-time}
\end{figure}

In the previous sections, we showed the effectiveness of learning, showing that \gls{ops} achieves the highest returns among the baselines. However, we have not addressed how \gls{ops} computationally benefits MARL when applied to multiple agents. To examine this, we have created \cref{fig:run-time}, which presents the median time for a timestep during training. It is clear that while \gls{ops} adds computational complexity over the fully shared networks, it scales significantly better than \gls{nps} does. In the BPS environments with 30 agents, \gls{ops} almost requires half the training time of \gls{nps} due to the substantially fewer trainable parameters. In BPS-h(3), training with \gls{nps} was infeasible since it requires 200 sets of parameters (50 more times than \gls{ops}). 

\subsection{Implementation Details}\label{sec:reproducibility}
In our experiments, we used Adam with learning rate of $3\mathrm{e}{-4}$, optimiser epsilon $1\mathrm{e}{-5}$, entropy coefficient $1\mathrm{e}{-2}$, and value, critic, and encoder-decoder networks with two layers of 64 or 128 units. Eight environments were sampled concurrently and 5-step returns were computed. The hyperparameters of entropy, learning rate, network width (but not depth), and n-step returns have been optimised using a coarse grid search, on one task per environment. A finer grid search on the entropy coefficient for the SMAC task was performed. For the encoder-decoder training, $m$ was set at 5, the KL loss was scaled by $1\mathrm{e}{-4}$, and we used batch size 128. For \cref{fig:colorprogression},  \gls{sap} and \gls{sapi} use two layers of 128 units, but \gls{sapi} (Scaled Up) uses two layers with a width of 128, 189, 236, 277, 313, 345, 375, and 401 for BPS-h with one to eight colors respectively. \Cref{fig:run-time} was generated in an AMD Epyc 7702 running Python 3 with environments sampled in parallel threads. 

\section{Related Work}
\textbf{Centralised Training with Decentralised Execution (CTDE): } A paradigm popular in cooperative MARL, assumes that during training all agents can access data from all other agents. After the training is completed, the agents stop having access to external data, and can only observe their own perspective of the environment. CTDE algorithms such as MADDPG~\cite{lowe_multi-agent_2017}, Q-MIX~\cite{rashid_qmix_2018}, and SEAC~\cite{christianos_shared_2020} all benefit from the centralised training stage and have been repeatedly shown to outperform non-CTDE baselines. \Gls{ops} also adheres to the CTDE paradigm and assumes that during training all information is shared.

\textbf{Parameter Sharing: } Sharing parameters between agents has a long history in MARL. \citet{tan_multi-agent_1993} investigates sharing policies between cooperative settings in non-deep RL settings. More recently, algorithms such as COMA~\cite{foerster_counterfactual_2018}, Q-Mix~\cite{rashid_qmix_2018}, or Mean Field RL~\cite{yang_mean_2018} share the parameters of neural networks similarly to our \gls{sap} and \gls{sapi} baselines. ROMA~\cite{wang_roma_2020} learns dynamic roles to share experience between agents that perform similar tasks. With \gls{ops} we do this operation statically in order to maximise computational efficiency, but we arrive at similar partitioning of agents in heterogenous SMAC tasks (\cref{fig:sampled_z}). The novelty of \gls{ops} does not come from sharing parameters, which is a well-established method in MARL, but that it creates neural network architectures in advance, allowing more efficient and effective sharing.

\textbf{Sharing Experience: } SEAC~\cite{christianos_shared_2020} shares experience between agents while maintaining separate policy and value networks. While SEAC achieves state-of-the-art performance, not only does it require one network per agent (i.e.\ \gls{nps}), it also stacks the experience of the agents leading to increased batch sizes. With \gls{ops} we forfeit the exploration benefits of SEAC but arrive at a method that may scale to hundreds of agents.

\textbf{Scaling MARL to more Agents: } Mean Field~\cite{yang_mean_2018} tackles MARL with numerous agents by approximating interactions between a single agent and the average effect of the population. While it is shown that convergence is improved, Mean Field RL shares parameters in a fashion similar to \gls{sap}. Our method operates as a pre-training step and attempts to find a network architecture configuration that improves learning. \Gls{ops} can be combined with MARL algorithms (centralised critic, value decomposition, or others) since it improves a different part of the RL procedure.

\section{Limitations and Future Work}

Partitioning the agents using samples collected before agents are allowed to learn a policy does come with a disadvantage. In situations where agents share dynamics and reward functions ($\hat{\cP^i}$ and $\hat{R^i}$) early in the policies' training but diverge later (e.g. agents are required to do the same task and then a different task in the same episode), learning the encoder-decoder with the initially collected samples may fail to properly partition agents. While in that case \gls{ops} will operate similarly to the full parameter sharing baselines like \gls{sap}, it could be further improved by regularly retraining the encoder-decoder model with newer experience and redistributing agents to clusters if they have diverged.

A more complicated situation arises when agents have identical dynamics and rewards but are meant to take on different roles. 
The benefit of sharing parameters (or not) in such a case is highly dependent on the nature of the specific environment.
While it may be possible that roles can be found and used to further partition agents if the \gls{ops} procedure is performed with trained policies (by recognising the difference in the sampling distributions), we leave such experiments and potential improvements to future work.

\section{Conclusion}
This paper explored existing methods for parameter sharing in MARL, identifying situations where they were ineffective. Our experiments suggested that sharing parameters indiscriminately between agents made learning harder since agents interfered with the learning of others (\cref{sec:exps:sap-i}). Therefore, we proposed a method for selective parameter sharing, that identified groups of agents that may benefit from sharing parameters. \Gls{ops} was shown to successfully recognise heterogeneous agents and assign them to different parameter sets, allowing MARL training to scale to hundreds of agents even when they were not homogeneous. Our method was shown to outperform other parameter sharing baselines in converged returns, and a non parameter sharing baseline both in converged returns and training speed. 


\section{Funding Disclosure}
This research was in part financially supported by the UK EPSRC Centre for Doctoral Training in Robotics and Autonomous Systems (F.C., G.P.), and the University of Edinburgh Enlightenment Scholarship (A.R.).

\bibliography{references,references_int}
\bibliographystyle{icml2021}
\end{document}